\documentclass[12pt]{article}
\usepackage{amsmath, amssymb, mathrsfs, mathtools, tensor, braket}
\usepackage{xcolor, graphicx, subfigure}
\graphicspath{{./_Figures/}}

\usepackage{cite, varioref}
\numberwithin{equation}{section}

\usepackage{enumitem, float}
\usepackage{caption}
\DeclareCaptionFormat{myformat}{#3}
\usepackage[compat=1.0.0]{tikz-feynman}

\evensidemargin \oddsidemargin

\begin{document}

\begin{titlepage}

\begin{center}
\vspace{5mm}
    
% Title
{\Large \bfseries
Effective Field Theory of Superconductivity
}\\[17mm]
% Authors
Yoonbai Kim,
~~SeungJun Jeon,
~~Hanwool Song
\\[3mm]  
% Departments & E-mail
{\itshape
Department of Physics,
Sungkyunkwan University,
Suwon 16419,
Korea
\\[-1mm]
yoonbai@skku.edu,~
sjjeon@skku.edu,~
hanwoolsong0@gmail.com
}
\end{center}
\vspace{15mm}

\begin{abstract}

A field theory of a Schr\"odinger type complex scalar field of Cooper pair, a $\text{U}(1)$ gauge field of electromagnetism, and a neutral scalar field of gapless acoustic phonon is proposed for superconductivity of $s$-waves.
Presence of the gapless neutral scalar field is justified as low energy residual acoustic phonon degrees in the context of effective field theory.
The critical coupling of quartic self-interaction of complex scalar field is computed from a $1$-loop level interaction balance between the repulsion mediated by massive degree of the $\text{U}(1)$ gauge field and the attraction mediated by massive Higgs degree, in the static limit.
The obtained net attraction or repulsion in perturbative regime matches the type I or I$\!$I superconductivity, respectively.
We find the new critical coupling of cubic Yukawa type interaction between the neutral and complex scalar fields from another tree level interaction balance between the Coulomb repulsion mediated by massless degree of the $\text{U}(1)$ gauge field and the attraction mediated by the gapless neutral scalar field, in the static limit.
Superconducting phase is realized at or in the vicinity of this critical coupling.
A huge discrepancy between the propagation speeds of photon and phonon gives a plausible explanation on low critical temperatures in conventional superconductors.

\end{abstract}

\end{titlepage}

\section{Introduction}

Though it took long time to develop the theory after experimental discovery \cite{onnes:1911}, conventional superconductivity of $s$-wave has two 
well-established standard theoretical descriptions. One is the fundamental Bardeen-Cooper-Schrieffer (BCS) theory \cite{Bardeen:1957mv, Cooper:1956zz} beginning with formation mechanism of the Cooper pairs \cite{Cooper:1956zz, Shankar:1993pf} from electrons and the other is the effective Ginzburg-Landau theory \cite{Ginzburg:1950sr, Abrikosov:1956sx}
based on the Ginzburg-Landau free energy of the order parameters, mostly time-independent. Despite its long history of several decades, the mother effective field theory of the Ginzburg-Landau theory in terms of time-dependent quantum fields is lacked until now. It has been seemed on the verge of being achieved for long time, but
we do not have any action in consensus for the effective field theory of superconductivity \cite{Nagaosa:1999ud, Wen:2004ym, Bennemann:2008, Coleman:2015, Arovas:2019}.

Then the period of high temperature superconductivity was initiated by breaking \cite{Bednorz:1986tc} the theoretically expected threshold, the maximum possible critical temperature around 20K$\sim$30K \cite{McMillan:1968pr, Allen:1968pr, Giustino:1968pr} in conventional superconductivity.
As the critical temperature passed the boiling temperature of liquid nitrogen \cite{Wu:1987te}, the subject was entering a boom era.
Recently superconductivity is a hot issue in relation with the superconductivity in twisted bilayer graphene \cite{Cao:2018xrk} and the superconductivity under high pressure \cite{Drozdov}.

Thus it is timely to find an effective field theory of conventional superconductivity including time dependence as a legitimate language for quantum dynamics.
In this paper we write an effective field theory of a Schr\"{o}dinger type complex scalar field of Cooper pair, a U(1) gauge field of electromagnetism, and a gapless neutral scalar field of acoustic phonon with electrically coupled constant background charge density for conventional superconductivity. First, we find the superconducting vacuum as the energetically favored vacuum configuration and test some required properties of it, e.g., we readily reproduce perfect conductivity and the Meissner effect.
We also discuss similarity and difference between the obtained energy for time-independent configurations and the Ginzburg-Landau free energy. Second, inclusion of the neutral scalar field for low energy gapless acoustic phonon is discussed and justified in the context of effective field theory 
\cite{Polchinski:1992ed}. About the critical coupling of  quartic self-interaction of complex scalar field for distinction of type I and I$\!$I superconductors 
\cite{Abrikosov:1956sx}, we confirm it as a balance of the repulsive interaction of the massive mode of $\text{U}(1)$ gauge field and the attractive interaction mediated by the massive Higgs degree of Cooper pairs. In the same manner, we discover the new critical coupling of cubic Yukawa type mutual interaction between Cooper pair and acoustic phonon from another static interaction balance of the Coulombic repulsion mediated by the massless degree of $\text{U}(1)$ gauge field and the long ranged attraction mediated by the gapless acoustic phonon. The perfect cancellation of the two interactions in tree level provides a plausible explanation for the curiosity on the absence or, at least, the feeble nature of the net interaction between two Cooper pairs in superconducting phase of arbitrary quartic self-interaction coupling despite of strong Coulombic repulsion. Since the obtained interaction balances are sustained in static force limit 
\cite{Abrikosov:1975a}, breakdown of them due to the largely different 
propagation speeds of relativistic photon and nonrelativistic matter fields may explain the low critical temperatures \cite{McMillan:1968pr, Allen:1968pr, Giustino:1968pr,Warlimonthandbook} in conventional superconductivity and may lead to possible temperature dependence of the critical value of quartic self-interaction coupling.

The rest of this paper is organized as follows. In section 2, we introduce the action and discuss the superconducting vacuum. In section 3, possible appearance of the gapless phonon field in effective field theory is discussed. In section 4 and 5, we derive two critical couplings by use of the interaction balances. We conclude in section 6 with discussions.

\section{Construction of Field Theory for Superconductivity and Superconducting Vacuum}

Conventional superconductivity has been described in terms of the 
Ginzburg-Landau free energy 
\cite{Ginzburg:1950sr, Arovas:2019, tinkham2004introduction, Bennemann:2008}. In the viewpoint of effective field theory including time evolution, this would-be effective field theory possesses at least the following fields.
Scalar potential $\Phi(t,x^{i})$ $(i= 1, 2, 3)$ and vector potential 
$A^{i}(t,x^{j})$ are combined as a U(1) gauge field $A^{\mu}=(\Phi/c,A^{i})$ ($\mu= t, 1, 2, 3$) with spacetime signature $(-,+,+,+)$, which describes electromagnetic waves propagating with the light speed $c$ in free space.
Its field strength tensor 
$F_{\mu\nu}=\partial_{\mu}A_{\nu}-\partial_{\nu}A_{\mu}$ with four-component spacetime derivatives $\partial_{t}=\partial/\partial t$ and 
$\partial_{i}=\partial/\partial x^{i}=(\boldsymbol{\nabla})_{i}$ combines electric and magnetic fields,
$(\boldsymbol{E})^{i} = c F_{i0}$ and $(\boldsymbol{B})^{i} = \frac{1}{2} \epsilon^{ijk} F_{jk}$.  
A complex scalar field $\Psi$ of mass $m=2m_{{\rm e}}$ and charge $q=-2e$ at tree level expresses a Cooper pair, which is neither a wavefunction in quantum mechanics nor an order parameter $
\Psi(t, x^{i})\sim
\braket{\psi_{\uparrow}(t,x^{i}) \psi_{\downarrow}(t,x^{i})}$ but a composite operator of two spin-up and down electrons in quantum field theory,
\begin{align}
\Psi(t,x^{i})\sim
\psi_{\uparrow}(t,x^{i}) \psi_{\downarrow}(t,x^{i})
.
\label{204}
\end{align}
Thus its compositeness is assumed to be hidden in the context of our description.
A gapless neutral scalar field $N$ describes a quantum field called the acoustic phonon originated from low frequency modes of lattice vibration with propagation speed $v_{N}$ which is presumably much slower than the light speed, $v_{N}\ll c$.  

Dynamics is governed by the action in thin or thick superconducting samples of flat slab shape,
\begingroup
\allowdisplaybreaks
\begin{align}
S 
=
& 
\int dt \int d^{2} \boldsymbol{x} \int dz \,
\Big[
- \frac{ \epsilon_{0} c^{2} }{4}
F_{\mu\nu} F^{\mu\nu}
+ \frac{i \hbar }{2} (\bar{\Psi} \mathcal{D}_{t} \Psi 
- \overline{\mathcal{D}_{t} \Psi} \Psi ) 
- \frac{\hbar^{2}}{2m} \overline{ \mathcal{D}_{i}\Psi}\mathcal{D}_{i}\Psi
\nonumber\\
&~~~~~~~~~~~~~~~~~~~~~~~~~
+ \frac{1}{2v_{N}^{2}}(\partial_{t}N)^{2}
-\frac{1}{2}(\partial_{i}N)^{2}
-V(|\Psi|,N) + q n_{\rm s} \Phi
\Big]
,
\label{201}
\end{align}
\endgroup
where nonrelativistic dynamics of
the complex scalar field $\Psi=|\Psi|e^{i\Omega}$ of amplitude $|\Psi|$ and phase $\Omega$ means a Schr\"{o}dinger type matter without characteristic speed, which couples minimally to electromagnetism via the gauge-covariant derivative
\begin{equation}
\mathcal{D}_{t} \Psi
=
\Big(
\frac{\partial}{\partial t}
+
i \frac{q}{ \hbar} \Phi
\Big) \Psi
,\qquad
\mathcal{D}_{i} \Psi
=
\Big(
\frac{\partial}{\partial x^{i}}
-
i \frac{q}{\hbar} A^{i}
\Big) \Psi
.
\end{equation} 
The last term in the action $S$ \eqref{201} expresses an electric coupling between the scalar potential $\Phi$ and a constant  matter density of 
background superconducting electrons $n_{\rm s}$ called the superfluid density.
Complex scalar field is self-interacting up to the quartic term of coupling constant $\lambda$ and mutually interacting with neutral scalar field through a cubic Yukawa type term of
coupling constant $g$.
Hence the potential of complex and neutral scalar fields is written by 
\begin{align}
V(|\Psi|,N) = \lambda (|\Psi|^{2} - v^{2}) \Big( |\Psi|^{2} + \frac{g}{\lambda} N - v^{2} \Big)
,
\label{206}
\end{align}
where $\lambda$ is positive quartic self-interaction coupling, $g$ is cubic Yukawa type coupling between neutral and complex scalar fields, and $v$ is vacuum 
expectation value.
Since the action \eqref{201} with the scalar potential \eqref{206} is invariant under sign flip of the neutral scalar field, $N \leftrightarrow -N$, with accompanied sign flip of the cubic Yukawa type coupling, $g\leftrightarrow -g$, it is enough to take into account nonpositive cubic Yukawa type coupling, $g \le 0$, as long as whole real values of the neutral scalar field are taken into account. 
Therefore, the proposed effective field theory is renormalizable by dimension counting, that can be established in perturbative quantum field theory. 

For a given action \eqref{201}, the Euler-Lagrange equations are read.
The charged complex scalar field
follows a gauged nonlinear Schr\"{o}dinger equation
\begin{equation}
i \hbar \mathcal{D}_{t}\Psi =
- \frac{\hbar^{2}}{2m} \mathcal{D}_{i}^{2}\Psi - 2 \lambda (v^{2}- |\Psi|^{2} )\Psi+ g N  \Psi
,\label{213}
\end{equation} 
and the neutral scalar field obeys an inhomogeneous linear wave equation for acoustic waves,
\begin{align}
\frac{1}{v_{N}^{2}}\frac{\partial^{2}N}{\partial t^{2}} - \boldsymbol{\nabla}^{2}N = - g (|\Psi|^{2} - v^{2})
. \label{214}
\end{align}
Classical dynamics of the \(\mathrm{U}(1)\) gauge field \(A^\mu\) is governed by an inhomogeneous Maxwell's equation named the Gauss' law
\begin{align}
\boldsymbol{\nabla} \cdot \boldsymbol{E} = \frac{\rho}{\epsilon_0}
,\label{209}
\end{align}
where $\rho$ in its right-hand side is charge density
\begin{align}
\rho = q(|\Psi|^{2} - n_{\rm s})
.\label{215}
\end{align}
The other inhomogeneous Maxwell equation is the Amp\'ere's law with 
displacement current term,
\begin{align}
- \frac{1}{c^{2}} \frac{\partial \boldsymbol{E}}{\partial t} 
+ \boldsymbol{\nabla} \times \boldsymbol{B} 
= \frac{\boldsymbol{j}}{\epsilon_{0} c^{2}}
,\label{222}
\end{align}
where the $\text{U}(1)$ current density is
\begin{align}
(\boldsymbol{j})^{i} = j^{i} = - i \frac{q\hbar}{2m} (\bar{\Psi} \mathcal{D}_{i} \Psi - \overline{\mathcal{D}_{i} \Psi} \Psi)
.\label{202}
\end{align}
The N\"{o}ther charge of the U(1) gauge symmetry is the electric charge%in SI unit system
\begin{align}
Q_{{\rm U}(1)} = \int d^{2}\boldsymbol{x}dz\,\rho 
\label{228}
\end{align}
whose finiteness requires localization of the charge density as a necessary condition,
\begin{align}
\lim_{|\boldsymbol{x}|\rightarrow\infty}|\Psi|^{2}=n_{{\rm s}}
.
\end{align}

\begin{figure}[h]
\centering
\includegraphics[width=0.55\textwidth]{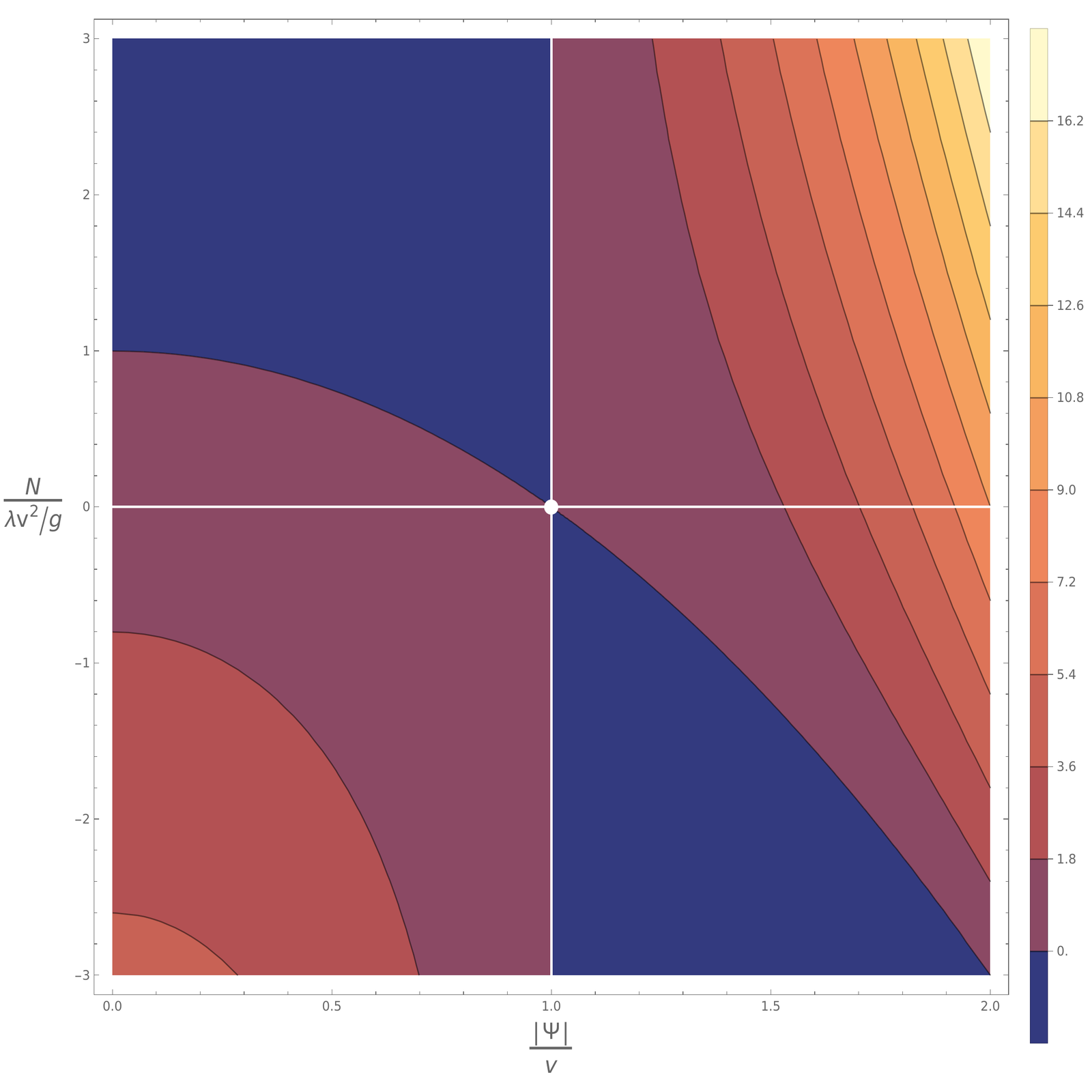}
\caption{Contour plot of the potential $V$ in which the white-colored dot at the center is $\braket{|\Psi|} = v$ and $\braket{N} = 0$. Horizontal axis denotes complex scalar amplitude $|\Psi|$ and vertical axis does neutral scalar field $N$.}
\label{fig:201}
\end{figure}

As shown in the contour map in figure \ref{fig:201}, the potential \eqref{206} does not include any stable minimum.
As every case of sound wave, the acoustic phonon described by the neutral scalar field has the minimum energy at equilibrium without displacement,
\begin{align}
N=0 .\label{311}
\end{align} Thus the only acceptable candidate of vacuum of zero energy is
\begin{align}
(\braket{|\Psi|}, \braket{N}) = (v,0) ,
\label{221}
\end{align}
which is stable along the scalar amplitude and is flat along the neutral scalar field.
Nonzero vacuum expectation value $\braket{|\Psi|} = v$ means spontaneous breakdown of the $\text{U}(1)$ symmetry.

In order to confirm the proposed vacuum \eqref{221} in classical level, the energy %examined
obtained from canonical energy-momentum tensor $T\indices{^\mu _\nu}$ is taken into account,
\begingroup
\allowdisplaybreaks
\begin{align}
E
&=  \int d^{2} \boldsymbol{x} \,dz(-T\indices{^t_t}) \nonumber \\
&=\int d^{2} \boldsymbol{x} \, dz \bigg\{\frac{\epsilon_{0}}{2} 
(\boldsymbol{E}^{2} + c^{2}\boldsymbol{B}^{2} ) 
+ \frac{\hbar^{2}}{2m}( \boldsymbol{\nabla} |\Psi| )^{2}+\frac{q^{2}}{2m} |\Psi|^{2}\Big( A^{i} - \frac{\hbar}{q} \partial_{i} \Omega \Big)^{2}\nonumber\\
&\, ~~~~~~~~~~~~~~~~~+ \frac{1}{2 v_{N}^{2}} (\partial_{t} N)^{2} 
+ \frac{1}{2} ( \boldsymbol{\nabla} N )^{2} + \lambda (|\Psi|^{2} - v^{2}) \Big( |\Psi|^{2} + \frac{g}{\lambda} N - v^{2} \Big)\nonumber\\
&\, ~~~~~~~~~~~~~~~~~- \epsilon_{0} \Phi \Big[ \boldsymbol{\nabla} \cdot \boldsymbol{E} - \frac{q}{\epsilon_0} (|\Psi|^{2} - n_{\rm s}) \Big]
+ \boldsymbol{\nabla} \cdot (\epsilon_{0}\boldsymbol{E}\Phi)
\bigg\}
.
\label{240}
\end{align}
\endgroup
The square bracket term in the last line of energy \eqref{240} vanishes by the Gauss' law \eqref{209} with charge density \eqref{215}.
Then, except the last total divergence term, all the six derivative terms of the energy \eqref{240} become positive semidefinite and the five derivative terms among six become zero by constant uniform scalar fields,
$\nabla|\Psi| = \partial_{t} N =\nabla N = 0 $,
and zero electric and magnetic fields,
$ \boldsymbol{E}=\boldsymbol{B}={\bf 0} $. 
Subsequently, the last total divergence term does not contribute to the energy by the zero electric field. With the help of zero vacuum configuration of the neutral scalar field \eqref{311}, the scalar potential \eqref{206} has zero
in the symmetry-broken vacuum of vacuum expectation value \eqref{221}, 
\begin{align}
|\Psi|=v \neq 0
, \label{207}
\end{align}
where this nonzero constancy of the complex scalar amplitude is called rigidity of the wavefunction.\footnote{This nonzero constant wavefunction is not normalizable in infinite space. Since the nonzero vacuum expectation value of the complex scalar field $\langle\Psi(t,x^{i})\rangle_{0}$ is not necessarily normalizable in this nonlinear field theory even though the complex scalar field, an operator, follows Schr\"{o}dinger type nonrelativistic dynamics. In this sense, the terminology ``rigidity of the wavefunction'' must be used with caution in the context of field theory.  }
 The vacuum configuration \eqref{207} achieved on the physical ground \eqref{311} coincides with the claimed vacuum \eqref{221} which involves a flat direction.
By substituting the aforementioned trivial vacuum configuration \eqref{221}, there remains only one term in the energy \eqref{240},
\begin{align}
E \ge \frac{q^{2} v^{2}}{2m} \int d^{2} \boldsymbol{x} \,dz \Big(A^{i}
-\frac{\hbar}{q}\partial_{i}\Omega\Big)^{2},
\end{align}
and minimum zero energy is achieved in terms of a pure gauge degree of freedom, a real function
$\Lambda(t,x^{i})$ obeying the single-valuedness condition 
$ \boldsymbol{\nabla} \times \boldsymbol{\nabla} \Lambda=0$, for the scalar phase
\begin{align}
\Omega=q\Lambda/\hbar
,\label{223}
\end{align}
and the gauge field, $\Phi=-\partial\Lambda/\partial t$ and 
$A^{i}=\partial_{i}\Lambda$.
If the symmetry-broken vacuum is electrically neutral without electric field $\boldsymbol{E}={\bf 0}$, zero charge density at any time and spatial point \eqref{215} determines the vacuum expectation value $v$, a theoretical 
parameter, in terms of the constant superfluid density  $n_{\rm s}$, a measurable physical quantity, as
\begin{align}
v^{2} = n_{\rm s}
.\label{305}
\end{align}
Once the vacuum expectation value is identified as the superfluid density \eqref{305}, the right-hand side of the Gauss' law \eqref{209} vanishes everywhere in the vacuum and hence the obtained vacuum is reconfirmed to be electrically neutral that naturally explains perfect conductivity.
In this vacuum of zero energy, the $\text{U}(1)$ current density \eqref{202} also trivially vanishes, $j^{i} = 0$, which is consistent with zero magnetic field through the Amp\'ere's law \eqref{222}.

The energy flux density is
\begin{align}
- T\indices{^i_t} = \epsilon_{0} c^{2} (\boldsymbol{E} \times \boldsymbol{B})^{i} - \frac{\hbar^2}{2m} (\overline{\mathcal{D}_{i} \Psi} \mathcal{D}_{t} \Psi + \overline{\mathcal{D}_{t} \Psi} \mathcal{D}_{i} \Psi) - \partial_{i} N \partial_{t} N 
,\label{219}
\end{align} 
and the momentum density is
\begin{align}
T\indices{^t_i} = \epsilon_{0} (\boldsymbol{E} \times \boldsymbol{B})^{i} + \frac{m}{q} j^{i} - \frac{1}{v_{N}^{2}} \partial_{t} N \partial_{i} N + q n_{\rm s} A^{i}
. \label{220}
\end{align}
The energy flux density and the momentum density are not symmetric, $-T\indices{^i_t} \neq c^{2} T\indices{^t_i}$, except for the Poynting vector part of electromagnetism since the matter part of Cooper pair and acoustic phonon possesses the Galilean boost symmetry instead of the Lorentz boost symmetry. In the symmetry-broken vacuum of type I superconducting state, the energy flux density
is zero, $-T\indices{^i_t} = 0$, and the momentum density
becomes a total spatial derivative term of the unphysical gauge degree of freedom, $T\indices{^t_i} = \partial_{i} (q n_{\rm s} \Lambda)$, in which the gauge degree of freedom can always be fixed to be zero, 
$\Lambda(t,\boldsymbol{x}) = 0$.
Stress components $T\indices{^i_j}$ of the energy-momentum tensor are symmetric $T\indices{^i_j} = T\indices{^j_i}$ with the help of the equations of motion due to the symmetry of spatial rotation,
\begingroup
\allowdisplaybreaks
\begin{align}
T\indices{^i_j} =&\, \Big[ \frac{\epsilon_0}{2} (\boldsymbol{E}^{2} 
+ c^{2}\boldsymbol{B}^{2}) + \frac{i \hbar}{2}(\bar{\Psi} \mathcal{D}_{t} \Psi - \overline{\mathcal{D}_{t} \Psi} \Psi ) 
- \frac{\hbar^{2}}{2m}|\mathcal{D}_{k}\Psi|^{2} 
+ \frac{1}{2v_{N}^{2}}(\partial_{t}N)^{2} -\frac{1}{2}(\partial_{k}N)^{2} \nonumber\\
&\,~~ - \lambda (|\Psi|^{2} - v^{2}) \Big( |\Psi|^{2} + \frac{g}{\lambda} N - v^{2} \Big) + q \Phi n_{\rm s} \Big] \delta_{ij} \nonumber\\
&\, - \epsilon_{0} E^{i} E^{j} - \epsilon_{0} c^{2} B^{i} B^{j} 
+ \frac{\hbar^{2}}{2m}(\overline{\mathcal{D}_{i} \Psi} \mathcal{D}_{j} \Psi + \overline{\mathcal{D}_{j} \Psi} \mathcal{D}_{i} \Psi) + \partial_{i} N \partial_{j} N
.
\end{align}
\endgroup
For the vacuum configuration, only the diagonal pressure components become total time derivative term of the unphysical gauge degree of freedom,
\begin{align}
T\indices{^i_j} = q n_{\rm s} \Phi \delta_{ij} = \frac{\partial}{\partial t} \big[ - q n_{\rm s} \Lambda (t,\boldsymbol{x}) \delta_{ij} \big]
,
\end{align}
which can always set to be zero by turning off the unphysical gauge degree of freedom, $\Lambda(t,\boldsymbol{x}) = 0$.

Since constant scalar phase is periodic, $ \Omega \in\mathbb{R} / 2\pi \mathbb{Z} $, and each vacuum 
$(|\Psi|, \Omega)=(v, \Omega)$ is superselective for every sample with 
macroscopic volume, the set of these superselective symmetry-broken vacuum configurations 
constitutes a cylinder ${\rm S}^{1} \times \mathbb{R}^{1}$ of radius $v$ as the vacuum manifold 
$\Omega$ $(0\le\Omega<2\pi)$, whose topology is characterized by the first homotopy of integer group, 
$\Pi_{1} (\text{S}^{1} \times \mathbb{R}^{1}) = \mathbb{Z}$.
The obtained configuration
of zero classical energy and zero winding number $n=0$ is the vacuum of minimum energy called the Higgs vacuum in relativistic field theories and the 
perfect superconducting state for conventional 
superconductors.
Topologically nontrivial sectors of nonzero winding numbers $n\ne0$ are dealt with the studies of charged vortices elsewhere~\cite{FTSC}.

In the presence of external magnetic field without external electric field, the $\text{U}(1)$ gauge field is decomposed into dynamical and background components,
\begin{align}
\Phi=\Phi+\Phi^{\rm ext},\qquad A_{i} = A_{i} + A^{\rm ext}_{i}.\label{501}
\end{align}
Let us consider the superconducting samples possessing a symmetry of spatial translation along the $z$ axis.
Since the condensed matter samples of interest have usually the shape of slab for studying superconductivity, spatial coordinates are divided into planar variables $\boldsymbol{x}=(x^{1},x^{2})=(x,y)$ on $xy$ plane and $x^{3} = z$ along the perpendicular direction for convenience.
If an external constant magnetic field perpendicular to the plane of the flat superconducting slab is applied, it is denoted by
\begin{align}
\boldsymbol{B}^{\rm ext}=(0,0,B^{\rm ext})
,\label{304}
\end{align}  
and the corresponding $\text{U}(1)$ gauge field \eqref{501} is written in the symmetric gauge,
\begin{align}
\Phi=\Phi,\qquad A_{i} = A_{i} + \frac{1}{2} \epsilon^{ij} x_{j} B^{\rm ext}
,~ (i=1,2), \qquad A_{3} = 0.
\label{224}
\end{align}
Substitution of the $\text{U}(1)$ gauge field \eqref{224} with the vacuum configuration of zero dynamical electric field of $\boldsymbol{E} = - \boldsymbol{\nabla} \Phi = \boldsymbol{0} $ and scalar fields \eqref{221} and \eqref{223} leaves some 
nonvanishing terms in the energy \eqref{240}
\begingroup
\allowdisplaybreaks
\begin{align}
E = \int d^{2} \boldsymbol{x} dz\, \bigg\{ \frac{\epsilon_{0} c^{2}}{2} \big[ B_{1}^{2} + B_{2}^{2} + (B - B^{\rm ext})^{2} \big] + \frac{q^{2} n_{\rm s}}{2m} \Big(A^{i} - \partial_{i} \Lambda + \frac{1}{2} \epsilon^{ij} x_{j} B^{\rm ext}\Big)^{2} \bigg\}
.
\label{230}
\end{align}
\endgroup
Then zero parallel magnetic field $B_{1} = B_{2} = 0$ reduces the energy \eqref{230} as
\begin{align}
E = \int d^{2} \boldsymbol{x} dz\, \bigg[ \frac{\epsilon_{0} c^{2}}{2} (B - B^{\rm ext})^{2} + \frac{q^{2} n_{\rm s}}{2m} (A^{i} - \partial_{i} \Lambda + \frac{1}{2} \epsilon^{ij} x_{j} B^{\rm ext})^{2} \bigg], \quad (i=1,2)
.
\end{align}
Up to the residual gauge degree of freedom $\Lambda$, minimum zero energy is recovered by canceling the external magnetic field, $B=B^{\rm ext}$ which is equivalent to $\displaystyle A^{i} = -\frac{1}{2} \epsilon^{ij} x_{j} B^{\rm ext}$.
Therefore, the net magnetic field should vanish in order to have the minimum zero energy irrespective of direction and magnitude of the external constant magnetic field \eqref{304}.
This energetically favored vacuum configuration which leads to the reproduction of zero energy explains perfect diamagnetism in conventional superconductivity \cite{tinkham2004introduction, Bennemann:2008, cohen2016fundamentals,Arovas:2019}. The Meissner effect is understood by the recovery of the superconducting vacuum of minimum zero energy through possible minimal change in response to the external 
magnetic field as was done in the Ginzburg-Landau free energy.

The massless neutral scalar field does not involve any characteristic scale and hence, two characteristic scales of the theory are identified:
One comes from the massive degree of U(1) gauge field and the other from the amplitude perturbation of complex scalar field with the nonzero vacuum expectation value.
In this symmetry-broken phase called the superconducting phase, the Amp\'{e}re's law including displacement term under the Lorenz gauge leads to a decoupled wave equation of the vector potential
with the help of no monopole condition and the Faraday's law,
\begin{align}
\frac{1}{c^{2}} \frac{\partial^{2} \boldsymbol{A}}{\partial t^{2}} 
- \boldsymbol{\nabla}^{2} \boldsymbol{A} 
= - \frac{q^{2} |\Psi|^{2}}{\epsilon_{0} c^{2} m} \Big( \boldsymbol{A} - \frac{\hbar}{q} \boldsymbol{\nabla} \Omega \Big) 
. \label{218}
\end{align}
If the scalar degrees are assumed to be frozen in the vacuum \eqref{221} and residual gauge degree of freedom is fixed by the unitary gauge also called the London gauge in case of constant phase $\Omega$,
\begin{align}
\bar{A}^{i}=A^{i}-\frac{\hbar}{q}\partial_{i}\Omega
,
\end{align} 
the wave equation of the vector potential absorbing the Goldstone degree $\Omega$ is linearized for topologically trivial sector,
\begingroup
\allowdisplaybreaks
\begin{align}
&\frac{1}{c^{2}} \frac{\partial^{2} \boldsymbol{\bar{A}}}{\partial t^{2}} - \boldsymbol{\nabla}^{2} \boldsymbol{\bar{A}} = - \frac{1}{\lambda_{\rm L}^{2}} \boldsymbol{\bar{A}} ,\label{227}
\end{align}
\endgroup
where $\lambda_{\rm L}$ is a characteristic length scale called the London penetration depth for magnetic fields in the superconducting phase,
\begin{align}
\lambda_{{\rm L}} = \sqrt{\frac{\epsilon_{0} m}{n_{\rm s}}} \frac{c}{|q|} .
\label{302}
\end{align}
For the Cooper pair at tree level, it has
\begin{align}
\lambda_{\rm L} = 3.76 \times 10^{6} / \sqrt{n_{\rm s}} 
~\text{m}^{-\frac{1}{2}}\sim 10^{-7}{\rm m}
,
\label{332}
\end{align}
where $n_{{\rm s}}\sim 10^{27}\text{m}^{-3}$ is used.
Since the light speed $c$ is the unique propagation speed of electromagnetic waves in free space, the corresponding time scale is $\lambda_{\rm L}/c \sim 10^{-16} \text{s}.$
Similarly, the scalar
equation \eqref{213} is linearized for the perturbed scalar amplitude 
$\delta|\Psi| $, defined by $ |\Psi| = v - \delta |\Psi|$, which is 
nonrelativistic version of the Higgs degree \cite{Anderson:1963pc},
\begin{align}
\Big( i \hbar \frac{\partial}{\partial t} + \frac{\hbar^{2}}{2m} 
\boldsymbol{\nabla}^{2} - 4\lambda v^{2} \Big) \delta |\Psi| = 0 .
\label{216}
\end{align}
From the linearized equation \eqref{216} for time-independent fields,
another characteristic length scale called the correlation length is read,
\begin{align}
\xi
=
\frac{\hbar}{2} \frac{1}{\sqrt{2 m\lambda n_{\rm s}}}
,\label{303}
\end{align}
where it measures the size of a Cooper pair and describes the distance between the two constituent electrons. According to the convention in condensed matter physics, the correlation length is not our 
$\xi$ \eqref{303} but $\sqrt{2}\xi$.\footnote{This unity looks more natural than $\sqrt{2}$ used in condensed matter community
in the viewpoint of equal mass of the Higgs and gauge boson, which is easily understood in the context of the Bogomolny limit~\cite{Bogomolny:1975de} and the supersymmetric Abelian Higgs model in relativistic regime 
\cite{DiVecchia:1977nxl, Witten:1978mh}.
The factor $\sqrt{2}$ comes from $1/2$ in front of quadratic spatial derivative term in the linearized scalar equation \eqref{216} and this $1/2$ is originated by taking nonrelativistic limit of the relativistic complex scalar field $\phi$ of mass $m$, 
$\phi = e^{-\frac{i}{\hbar} m c^{2} t} \Psi$.
}
Its formula for the Cooper pair is 
\begin{align}
\xi = \frac{2.76 \times 10^{14}}{\sqrt{\lambda n_{\rm s}}} 
~ \text{kg}^{\frac{1}{2}}\, \text{m}^{2}\, \text{s}^{-1}
.
\label{333}
\end{align}
By substituting the measured values of London penetration depth $\lambda_{{\rm L}}\simeq(0.5\sim5)\times10^{-7}{\rm m}$ and correlation length $\xi$ of the order of $10^{-6}{\rm m}$ which may say $10^{-7}{\rm m}<\xi<10^{-5}{\rm m}$ \cite{kittelintroduction, cohen2016fundamentals, Warlimonthandbook}, we read in reverse the value of the constant density of superconducting electrons for a Cooper pair, $n_{{\rm s}}\simeq 10^{27}{\rm m}^{-3}$, and allowed range of quartic self-interaction coupling, $\lambda\sim10^{30}\sim10^{34}{\rm eVm}^{3}$.

Though the superconducting vacuum configuration \eqref{221} is obtained with the help of physically reasonable equilibrium of acoustic phonon, its instability is expected as shown in the contour plot of scalar potential in figure \ref{fig:201}. Therefore, stabilization of the phonon seems to be taken into account additionally but it is beyond the scope of the current work.

\section{Effective Field Theory with Phonon}

When the neutral scalar field is turned off, $N=0$, the energy for static configurations \eqref{240} without the Gauss' law and total divergence term is formally equal to the Ginzburg-Landau free energy for conventional superconductors of $s$-wave \cite{Ginzburg:1950sr,tinkham2004introduction, Arovas:2019, Bennemann:2008}.
In other words, field contents of our field theoretic model \eqref{201} in its static limit is distinguished from the Ginzburg-Landau theory by the gapless neutral scalar field $N$ depicting an acoustic phonon.
At first glance, existence of the neutral scalar field in this effective field theory looks awkward because the phonon has played a role of combining two electrons to form a Cooper pair as understood in the BCS theory \cite{Bardeen:1955, Giustino:1968pr, Polchinski:1992ed, Shankar:1993pf}.

In the framework of effective field theory, we begin with the action with cubic Yukawa type interaction including vacuum expectation value but without the quartic self-interaction between Cooper pairs at classical level,
\begingroup
\allowdisplaybreaks
\begin{align}
S_{0}=
&\, \int d t d^{2} \boldsymbol{x} dz\, \bigg[ -\frac{\epsilon_{0} c^{2}}{4} F_{\mu\nu} F^{\mu\nu} + \frac{i\hbar}{2} (\bar{\Psi} \mathcal{D}_{t} \Psi - \overline{\mathcal{D}_{t} \Psi}\Psi) 
- \frac{\hbar^{2}}{2m} \overline{\mathcal{D}_{i} \Psi} \mathcal{D}_{i} \Psi + q n_{\rm s} \Phi \nonumber\\
&\, ~~~~~~~~~~~~~~~~~ + \frac{1}{2v_{N}^{2}} (\partial_{t} N)^{2} 
- \frac{1}{2} (\partial_{i} N)^{2}- g N (|\Psi|^{2} - v^{2})
\bigg] 
.
\label{331}
\end{align}
\endgroup
If the linear neutral scalar field is integrated out, we have a formal 
expression of an effective action,
\begingroup
\allowdisplaybreaks
\begin{align}
S_{0\,\rm eff} =&\, \int d t d^{2} \boldsymbol{x} dz\, \bigg[ 
-\frac{\epsilon_{0} c^{2}}{4} F_{\mu\nu} F^{\mu\nu} 
+ \frac{i\hbar}{2} (\bar{\Psi} \mathcal{D}_{t} \Psi 
- \overline{\mathcal{D}_{t} \Psi}\Psi) - \frac{\hbar^{2}}{2m} 
\overline{\mathcal{D}_{i} \Psi} \mathcal{D}_{i} \Psi 
+ q n_{\rm s} \Phi \nonumber\\
&\,~~~~~~~~~~~~~~~~~ + \frac{g^{2}}{2} (|\Psi|^{2} - v^{2}) 
\Big(\frac{1}{v_{N}^{2}} \frac{\partial^{2}}{ \partial t^{2}} - \partial_{i}^{2} \Big)^{-1} (|\Psi|^{2} - v^{2}) \bigg] \nonumber\\
&\, + \frac{i}{2} \ln \bigg[ \det\Big( \frac{1}{v_{N}^{2}} \partial_{t}^{2} - \partial_{i}^{2} \Big) \bigg]
.
\label{328}
\end{align}
\endgroup
In the limit of high frequencies of the phonon field, high cutoff frequency is introduced \cite{Burgess:2020tbq},
\begin{align}
\omega_{{\rm H}}\gg v_{N}|\boldsymbol{k}_{{\rm H}}|
.
\label{329}
\end{align} 
Then the derivatives can crudely be approximated by 
$[(1/v_{N}^{2})(\partial/\partial t)^{2}-\partial_{i}^{2}]^{-1} \approx - v_{N}^{2} / \omega_{{\rm H}}^{2}$ and so can be the effective action \eqref{328} up to redundant constant,
\begingroup
\allowdisplaybreaks
\begin{align}
S_{0\,\rm eff} \approx&\, \int d t d^{2} \boldsymbol{x} dz\, \bigg[ 
-\frac{\epsilon_{0} c^{2}}{4} F_{\mu\nu} F^{\mu\nu} 
+ \frac{i\hbar}{2} (\bar{\Psi} \mathcal{D}_{t} \Psi 
- \overline{\mathcal{D}_{t} \Psi}\Psi) - \frac{\hbar^{2}}{2m} 
\overline{\mathcal{D}_{i} \Psi} \mathcal{D}_{i} \Psi + q n_{\rm s} \Phi 
\nonumber\\
&\,~~~~~~~~~~~~~~~~~ - \frac{g^{2} v_{N}^{2}}{2 \omega_{{\rm H}}^{2}} (|\Psi|^{2} - v^{2})^{2}  \bigg]
.
\label{330}
\end{align}
\endgroup
The obtained effective action \eqref{330} looks to play the role of mother field theory including low energy dynamics of the Cooper pairs and its energy of time-independent configurations proceeds toward the Ginzburg-Landau free energy.

In the viewpoint of effective field theory,
complete integration out of the acoustic phonon by use of its linearity in the formal expression \eqref{328} and the crude approximation for high frequency modes \eqref{329} are incompatible. Since the gapless acoustic phonon is not an auxiliary field but a dynamical field, it always includes low frequency modes in the condensed matter samples of macroscopic sizes and hence low and high frequency modes should be divided 
\cite{Shankar:1993pf, Polchinski:1992ed},
\begin{align}
N~\rightarrow~N_{{\rm L}}+N_{{\rm H}}
,
\end{align}  
where the path integral measure is also assumed to be divided as $[dN]=[dN_{{\rm L}}][dN_{{\rm H}}]$ in accordance with linearity of the phonon. Integration of the high frequency modes of the acoustic phonon $N_{{\rm H}}$ leads to another formal expression of the effective action
\begingroup
\allowdisplaybreaks
\begin{align}
S_{\rm eff} =&\, \int d t d^{2} \boldsymbol{x} dz\, \bigg[ -\frac{\epsilon_{0} c^{2}}{4} F_{\mu\nu} F^{\mu\nu} + \frac{i\hbar}{2} (\bar{\Psi} \mathcal{D}_{t} \Psi - \overline{\mathcal{D}_{t} \Psi} \Psi) 
- \frac{\hbar^{2}}{2m} \overline{\mathcal{D}_{i} \Psi} \mathcal{D}_{i} \Psi + q n_{\rm s} \Phi\nonumber\\
&\, ~~~~~~~~~~~~~~~~~ + \frac{1}{2v_{N}^{2}} (\partial_{t} N_{{\rm L}})^{2} - \frac{1}{2} (\partial_{i} N_{{\rm L}})^{2} - \frac{g^{2} v_{N}^{2}}{2\omega_{\rm H}^{2}} (|\Psi|^{2} - v^{2})^{2} - g N_{{\rm L}} (|\Psi|^{2} - v^{2})  \bigg] \nonumber\\
&\, + \frac{i}{2} \ln \bigg[ \det\Big( \frac{1}{v_{N}^{2}} \partial_{t}^{2} - \partial_{i}^{2} \Big) \bigg]
.
\label{362}
\end{align}
\endgroup
Since the high frequency approximation \eqref{362} is consistent with the integrated high frequency phonon modes, it is approximated to become nothing but the proposed action \eqref{201} possessing the scalar potential \eqref{206} with the help of replacement $N_{{\rm L}}\rightarrow N$.
After removal of redundant constant vacuum contribution even if it is infinite, comparison of the two actions \eqref{201} and \eqref{362} gives the identification of quartic self-interaction coupling in terms of the high cutoff frequency, \begin{align} 
\lambda=\frac{g^{2} v_{N}^{2}}{2\omega_{{\rm H}}^{2} }
.
\label{348}
\end{align}
 
\begin{figure}[h]
\centering
\includegraphics[width=0.65\textwidth]{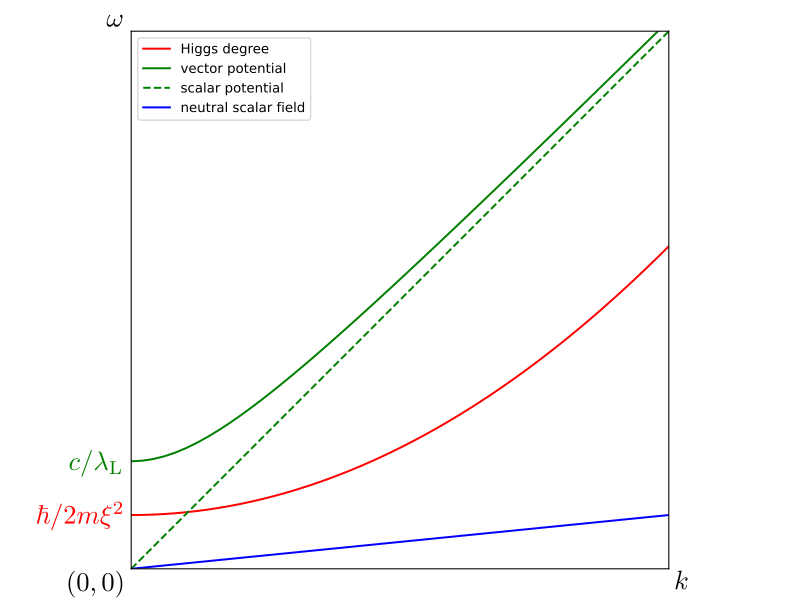}
\caption{Dispersion relations of the vector potential (the green-colored hyperbolic curve of vertical intercept $c/\lambda_{\rm L}$), the scalar potential (the green-colored dashed conic line), the Higgs degree (the red-colored parabolic curve of vertical intercept $\hbar/2m\xi^2$), and the neutral scalar field (the blue-colored conic line).}
\label{fig:301}
\end{figure}

It is turn to discuss possible high cutoff frequency in conventional superconductivity. Above all we examine the equations of motion and read the dispersion relations of the excited fields in order to read the energy gaps.
The dispersion relation for the Higgs degree $\delta |\Psi|$, neutral perturbative excitation from the vacuum \eqref{221}, is parabolic,
\begin{align}
\omega=\frac{\hbar}{2m} \Big[ (\boldsymbol{k}^{2}+k_{z}^{2}) + \frac{1}{\xi^2} \Big]
.\label{307}
\end{align}
 As illustrated by the red-colored curve in figure \ref{fig:301}, the energy gap of the Higgs degree is 
\begin{align}
\Lambda_{\xi}=\frac{\hbar^{2}}{2m \xi^{2}} = 4\lambda n_{\rm s}
.
\label{335}
\end{align}
The dispersion relation of the gauge-fixed vector potential $\boldsymbol{\bar{A}}$ read from \eqref{227} is hyperbolic,
\begin{align}
\omega^{2}=c^{2} \Big[ (\boldsymbol{k}^{2}+k_{z}^{2}) + \frac{1}{\lambda_{\rm L}^2} \Big]
,\label{306}
\end{align}
where $\omega$ is angular frequency and $(\boldsymbol{k},k_{z})$ wave vector.
It implies a massive degree of electromagnetic waves in the superconducting phase. As illustrated by the green-colored solid curve in figure \ref{fig:301}, the energy gap of electromagnetic field is
\begin{align}
\Lambda_{{\rm L}}=\frac{\hbar c}{\lambda_{\rm L}}
.
\label{336}
\end{align} With the help of the condition \eqref{305} for the superconducting vacuum, the equation of scalar potential $\Phi$ becomes homogeneous and its dispersion relation is unaltered from massless electromagnetic waves,
\begin{align}
\omega^{2} = c^{2} (\boldsymbol{k}^{2} + k_{z}^{2})
,
\label{308}
\end{align}
which defines an asymptotic cone of the hyperboloid as illustrated by the green-colored dashed line in figure \ref{fig:301}.
Similarly, in the superconducting vacuum \eqref{221} obeying the condition \eqref{305}, the neutral scalar field equation \eqref{214} reduces to a homogeneous linear wave equation
\begin{align}
\frac{1}{v_{N}^{2}} \frac{\partial^{2} N}{\partial t^{2}} - \boldsymbol{\nabla}^{2} N = 0, 
\end{align}
and
the dispersion relation forms a cone of wide angle for a typical gapless acoustic wave of propagation speed $v_{N}$ in nonrelativistic regime, $v_{N}\ll c$, as illustrated by the blue-colored straight line in figure \ref{fig:301},
\begin{align}
\omega^{2}=v_{{N}}^{2}(\boldsymbol{k}^{2}+k_{z}^{2})
.\label{309}
\end{align}
If a condensed matter sample of conventional superconductivity is taken into account, the IR cutoff $\Lambda_{{\rm IR}}$ for
gapless acoustic phonon of the dispersion relation \eqref{309} can naturally be chosen by the energy scale of the longest wavelength of the order of the size $L$ of the given sample, e.g., $L \sim$ a few millimeters,
which is estimated as $\Lambda_{{\rm IR}}=\hbar\omega_{L}=2\pi\hbar v_{N}/L \sim 3.92 \times 10^{-8} {\rm eV}$.
Since the UV cut for the highest frequency of phonon is given by the Debye frequency $\omega_{{\rm D}}$ of the Debye energy, $\Lambda_{{\rm D}}=\hbar\omega_{{\rm D}} \sim 0.025 \text{eV}$, the corresponding number of modes for the residual acoustic phonon $N=N_{{\rm L}}$ is huge,
\begin{align}
n_{N}=\frac{\Lambda_{{\rm D}}}{\Lambda_{{\rm IR}}}
 \sim 2.20 \times 10^{5 \sim 6}
.
\label{347}
\end{align}
Even for the lower energy scale of the critical temperature of typical superconductors $T_{{\rm c}}$, $\Lambda_{T_{{\rm c}}}\sim k_{{\rm B}}T_{{\rm c}} \sim 3.45 \times10^{-4}{\rm eV}$, $n_{N}$ is about the order of $10^{4}$. Therefore, the residual neutral scalar field of gapless acoustic phonon $N=N_{{\rm L}}$ carrying the energy lower than $\hbar\omega_{{\rm H}}$ is likely to survive in the low energy effective field theory as in the proposed action \eqref{201} after integrating out the high frequency modes of the phonon field $N_{{\rm H}}$.

If we compute the energy for time-independent configurations, it is read from the energy \eqref{240} by turning off the neutral scalar field of acoustic phonon and substituting the quartic self-interaction coupling $\lambda$ in terms of the cutoff frequency $\omega_{{\rm H}}$ \eqref{348}.
Since the vacuum configuration \eqref{207} including vanishing electric field fulfills trivially the Gauss' law, the effective field theory of $S_{{\rm eff}}$ \eqref{362} can be the first candidate of mother quantum field theory of the Ginzburg-Landau theory.
Resemblance and discrepancy between the Ginzburg-Landau theory whose Ginzburg-Landau free energy is free from the Gauss' law, and the quantum field theory of the effective action $S_{{\rm eff}}$ \eqref{362}, constrained by the Gauss' law \eqref{209}, will be discussed elsewhere in relation with characters of the vortices which follow the Gauss' law nontrivially \cite{FTSC}.
Until now a field-theoretic origin of the Ginzburg-Landau free energy seems to be identified as the energy not from the effective action $S_{{\rm eff}}$ \eqref{362} but from the effective action $S_{0\,{\rm eff}}$ \eqref{330} without the unidentifiable adjective ``free''.

\section{Borderline of Type I and I$\!$I Superconductivity as Interaction Balance}

The quantum theory of our interest possesses two length scales, the correlation length $\xi$ \eqref{303} and the London penetration depth $\lambda_{\rm L}$ \eqref{302}, and their ratio is known to be the Ginzburg-Landau parameter
\begin{align}
\kappa = \frac{\lambda_{\rm L}}{\xi} 
= \frac{2mc \sqrt{2\epsilon_{0} \lambda}}{\hbar |q|}  
\equiv \sqrt{\frac{\lambda}{\lambda_{\rm c}}}
.
\label{349}
\end{align}
When the two length scales are equal, the Ginzburg-Landau parameter becomes unity
\begin{align}
\kappa = \frac{\lambda_{\rm L}}{\xi} = 1
,\label{314}
\end{align}
and subsequently the critical value of the quartic self-interaction coupling is read,
\begin{align}
\lambda=
\lambda_{\rm c} = \frac{\hbar^{2} q^{2}}{8\epsilon_{0} m^{2} c^{2}} 
.
\label{312}
\end{align}
For the Cooper pair at tree level, its value is $\lambda_{\rm c} = 2.14 \times10^{-52} ~\text{kg}\, \text{m}^{5}\, \text{s}^{-2}$.
In conventional superconductivity, the critical coupling \eqref{312} describes the borderline of type I and I$\!$I superconductivity \cite{Abrikosov:1956sx}. A natural question is to figure out the reason why this borderline appears in the critical value \eqref{312} at least in perturbative regime of the effective field theory of action \eqref{201}.
In this section, we show that it can be understood by the balance of static forces between two Cooper pairs except the repulsive contact interaction by the quartic self-interacting vertex and the attractive short-ranged interaction by its cubic self-interacting vertex of complex scalar field in the superconducting phase.

Since the London penetration depth \eqref{302} comes from the massive degree of the U(1) gauge field expressed by the vector potential $A^{i}$, the propagator of the vector potential part of the gauge boson $A^\mu = (\Phi/c, A^{i})$ is taken into account with the help of the $R_{\xi}$ gauge fixing condition,\begingroup
\allowdisplaybreaks
\begin{align}
&\Delta^{ij} (\omega, \boldsymbol{k}) 
= \frac{\delta^{ij} - \frac{k^{i} k^{j}}{\boldsymbol{k}^{2}}}{\hbar^{2} \omega^{2} - c^{2} \Big[ (\boldsymbol{p}^{2} + p_{z}^{2}) + \frac{\hbar^{2}}{\lambda_{\rm L}^{2}} \Big] + i \epsilon} + \xi_{g} \frac{\frac{k^{i} k^{j}}{\boldsymbol{k}^{2}}}{\xi_{g} \hbar^{2} \omega^{2} - c^{2} \Big[ (\boldsymbol{p}^{2} + p_{z}^{2}) + \xi_{g} \frac{\hbar^{2}}{\lambda_{\rm L}^{2}} \Big] + i \epsilon} 
,\label{317}
\end{align}
\endgroup
where $\boldsymbol{p}^{2} + p_{z}^{2} 
= \hbar^{2} (\boldsymbol{k}^{2} + k_{z}^{2})$
and
$\xi_{g}$ is a parameter of the $R_{\xi}$-gauge. The propagator of amplitude of the complex scalar field about the symmetry-broken vacuum is automatically read
\begin{align}
\Delta_{\delta |\Psi|} (\omega, \boldsymbol{k}) = \frac{1}{\hbar \omega - \frac{1}{2m} \Big[ (\boldsymbol{p}^{2} + p_{z}^{2}) + \frac{\hbar^{2}}{\xi^{2}} \Big] + i \epsilon } .
\label{315}
\end{align}
For calculation of off-shell contributions on the superconducting vacuum, one quartic vertex for the scalar self-interaction is $-i\lambda$ and the other quartic vertex among two complex scalar fields and two vector potential parts of the U(1) gauge field, $-iq^{2}/2m$, is derived from the spatial components of quadratic covariant derivative term in the action \eqref{201}.

Leading order correction to the interaction between two Cooper pairs comes from the 1-loop contributions for which the two Feynman diagrams are in figure below:

\begin{center}
\begin{tikzpicture}
\begin{feynman}[large]
\vertex (i1);
\vertex [below right=of i1] (a);
\vertex [below left=of a] (i2);
\vertex [right=of a] (b);
\vertex [above right=of b] (f1);
\vertex [below right=of b] (f2);

\diagram* {
(i2) -- [fermion] (a) -- [fermion] (i1),
(a) -- [photon, quarter left] (b) -- [photon, quarter left] (a),
(f2) -- [fermion] (b) -- [fermion] (f1),
};
\end{feynman}
\end{tikzpicture}
\hspace{8em}
\begin{tikzpicture}
\begin{feynman}[large]
\vertex (i1);
\vertex [below right=of i1] (a);
\vertex [below left=of a] (i2);
\vertex [right=of a] (b);
\vertex [above right=of b] (f1);
\vertex [below right=of b] (f2);

\diagram* {
(i2) -- [fermion] (a) -- [fermion] (i1),
(a) -- [fermion, quarter left] (b) -- [fermion, quarter left] (a),
(f2) -- [fermion] (b) -- [fermion] (f1),
};
\end{feynman}
\end{tikzpicture}
\end{center}

%\begin{figure}[H]
%\centering
%\subfigure{\includegraphics[width=0.3\textwidth]{feynp4g2.png}}
%\qquad
%\subfigure{\includegraphics[width=0.3\textwidth]{feynp4p2.png}}
%\end{figure}

\noindent
Correction to the vertex function is made by the first photon loop,
\begin{align}
\frac{- \frac{q^{4}}{\epsilon_{0}^{2} m^{2} c^{4}}}{\Big( 
\frac{\omega^{\prime 2}}{c^{2}} - \boldsymbol{k}^{\prime 2} -k_{z}^{\prime 2}- \frac{1}{\lambda_{\rm L}^{2}}\Big) \Big( \frac{\omega^{2}}{c^{2}} - \boldsymbol{k}^{2} -k_{z}^{2}- \frac{1}{\lambda_{\rm L}^{2}}\Big)}
,
\label{360}
\end{align} 
and by the second Higgs loop in the quadratic order $\mathcal{O}(\lambda^{2})$,
\begin{align}
\frac{\frac{64 m^{2} \lambda^{2}}{\hbar^{4}}}{\Big(\frac{2m \omega^{\prime}}{\hbar} - \boldsymbol{k}^{\prime 2} -k_{z}^{\prime 2}- \frac{1}{\xi^{2}}\Big) \Big(\frac{2m \omega}{\hbar} - \boldsymbol{k}^{2} -k_{z}^{2} - \frac{1}{\xi^{2}}\Big)}
.
\label{361}
\end{align}
If the two energy gaps in the denominators of both vertex functions \eqref{360}--\eqref{361} are 
estimated by use of the measured values of London penetration depth 
$\lambda_{{\rm L}}$ and correlation length $\xi$ \cite{Warlimonthandbook},
\begingroup
\allowdisplaybreaks
\begin{align}
&\frac{\hbar c}{\lambda_{\rm L}}
=(0.38 \sim 3.8)\times10^{-1}{\rm eV}
,
\label{334}
\\
&\frac{\hbar^{2}}{2m \xi^{2}} 
\sim 0.27 \times 10^{-1}{\rm eV}
,
\label{338}
\end{align}
\endgroup  
they are comparable to the energy scale of Debye frequency $\omega_{{\rm D}}$, $\hbar\omega_{{\rm D}}\sim2.5\times10^{-2}{\rm eV}$, but higher than that from the typical  critical temperature $T_{{\rm c}}$, $k_{{\rm B}}T_{{\rm c}}\sim10^{-4}{\rm eV}$. Hence the frequency parts, $\omega$ and $\omega^{\prime}$, in the denominators of both vertex functions \eqref{360}--\eqref{361} can be suppressed and seem to be negligible in low energy dynamics. Under this static approximation, the origin of the interaction by the U(1) gauge field loop becomes almost magnetic, that is consistent with the equation for time-independent vector potential $A^{i}(\boldsymbol{x}, z)$ with the energy gap \eqref{227} derived from the Amp\'{e}re's law \eqref{222} assisted by the Lorenz gauge.  

If the quartic self-interaction coupling has the critical value 
$\lambda=\lambda_{{\rm c}}$ \eqref{312}, both the length scales in the denominators become equal, $\lambda_{\rm L} = \sqrt{\frac{\epsilon_{0} m}{n_{\rm s}}} \frac{c}{|q|} = \xi$, and absolute values of the numerators are equal, $\frac{64 m^{2} \lambda_{{\rm c}}^{2}}{\hbar^{4}}=\frac{q^{4}}{\epsilon_{0}^{2} m^{2} c^{4}}$, simultaneously. Since the overall sign is opposite, two 1-loop contributions in \eqref{360} and \eqref{361} are exactly cancelled for the critical quartic self-interaction coupling in the static limit.
This cancellation means that the repulsion mediated by virtual massive photon degree is perfectly cancelled by the attraction mediated by virtual massive Higgs degree. Both strengths of the interactions expressed in the numerators are the same and both length scales of the interactions in the denominators are equal as also in \eqref{314}, and thus the net interaction from these two 1-loop corrections in static limit is zero everywhere throughout the superconducting sample with critical quartic self-interaction coupling. The 1-loop calculation shows that the borderline between type I and I$\!$I superconductors exists at the critical coupling \eqref{312} as predicted first by use of the Ginzburg-Landau free energy \cite{Abrikosov:1956sx} and is understood as an exact cancellation of the attractive and repulsive static short-ranged interactions in the context of perturbative quantum field theory.
For weak quartic self-interaction coupling $0\le \lambda/\lambda_{\rm c} <1$ of equivalently small Ginzburg-Landau parameter $\kappa < 1$, the strength of attractive interaction becomes weaker but the correlation length $\xi$ \eqref{303} is longer than the London penetration depth $\lambda_{\rm L}$ \eqref{302}.
For this type I superconductivity, the net short ranged interaction is repulsive in shorter distance but turns to be attractive at relatively larger distance. For strong quartic self-interaction coupling $\lambda/\lambda_{\rm c} >1$ of equivalently large Ginzburg-Landau parameter $\kappa > 1$, the strength of attractive interaction becomes stronger but the correlation length $\xi$ \eqref{303} is shorter than the London penetration depth $\lambda_{\rm L}$ \eqref{302}. For this type I$\!$I superconductivity,
the net short ranged interaction is attractive in shorter distance but turns to be repulsive at relatively larger distance.

\section{Critical Phonon Coupling as Condition for Superconductivity}
         
The effective action \eqref{201} of our interest involves one more coupling $g$ from a cubic Yukawa type mutual interaction term in addition to the electromagnetic coupling $q/\sqrt{\epsilon_{0}}$ and the coupling $\lambda$ of a quartic self-interaction of complex scalar field. Let us explore the role of this cubic Yukawa type coupling in the viewpoint of interaction balance. A naive but fundamental question in any effective theory of superconductivity including the complex order parameter of Cooper pair is:
How can two Cooper pairs of negative electric charge $q=-2e$ be at least almost noninteracting in a superconducting sample despite of the Coulomb repulsion between them? A possible answer on this question seems electrical neutrality of the Higgs degree in the Ginzburg-Landau theory and in the Abelian Higgs model.
We address this question in the context of the proposed effective field theory of Schr\"odinger type complex scalar field by discussing possible cancellation of the Coulomb force between two charged Cooper pairs in a superconducting sample from here on.

Under the Lorenz gauge, the Gauss' law \eqref{209} leads to the decoupled wave equation of the scalar potential
with the help of no monopole condition and the Faraday's law,
\begin{align}
\frac{1}{c^{2}} \frac{\partial^{2} \Phi}{\partial t^{2}} - \boldsymbol{\nabla}^{2} \Phi = \frac{q}{\epsilon_{0}} (|\Psi|^{2} - n_{\rm s}) . \label{217}
\end{align}
Once the vacuum expectation value is identified as the constant superfluid density \eqref{305}, the right-hand sides of the Gauss' law \eqref{217} and the neutral scalar field equation \eqref{214} become identical up to a proportionality constant even with arbitrary complex scalar amplitude $|\Psi|$.
For time-independent configurations, the scalar potential is linearly related to the neutral scalar field,
\begin{align}
\Phi = - \frac{q}{\epsilon_{0}g} N + f, \label{310}
\end{align}
where $f=f(\boldsymbol{x})$ is an arbitrary time-independent function whose Laplacian vanishes, $\nabla^{2} f=0$. The undetermined scalar function $f$ is nothing but the residual gauge degree of freedom under the Coulomb gauge, which can always set to be zero. Then the linear proportionality from \eqref{310} implies possible cancellation of two static forces from the U(1) gauge field and the gapless acoustic phonon at classical level.

For tree level calculation, we read the propagator of the scalar potential part expressing massless degree
of the gauge boson $A^\mu = (\Phi/c, A^{i})$ under the $R_{\xi}$ gauge fixing condition,
\begin{align}
\Delta_{\Phi} (\omega, \boldsymbol{k}) = \frac{1}{\hbar^{2} \omega^{2} - \xi_{g} c^{2} (\boldsymbol{p}^{2} + p_{z}^{2}) + i \epsilon}
,\label{316}
\end{align}
and that of gapless acoustic phonon
\begin{align}
\Delta_{N} (\omega, \boldsymbol{k}) = \frac{1}{\hbar^{2} \omega^{2} - v_{N}^{2} (\boldsymbol{p}^{2} + p_{z}^{2}) + i \epsilon}
.\label{322}
\end{align} 
In the superconducting phase, one cubic interaction vertex of a gauge field and two complex scalar fields comes only from the time component of linear covariant derivative term and has $q/\sqrt{\epsilon_{0}}$. The other cubic interaction vertex of a neutral scalar field and two complex scalar fields is derived from the cubic Yukawa type interaction term and has $g$. Note that two linear contact terms $qn_{{\rm s}}\Phi$ and $gv^{2}N$ between constant background density and a field in the action \eqref{201} with the potential $V$ \eqref{206} disappear in the symmetry-broken superconducting phase by the cancellation due to the constraint \eqref{305} derived from the condition for the vacuum without electric field.

In the framework of perturbative quantum field theory, scattering amplitude of two Cooper pairs is considered in the symmetry-broken phase and the leading interactions are given by the two terms of quadratic order except the aforementioned short-ranged attractive self-interaction of the order $\mathcal{O}(\lambda^{2})$ mediated by a single massive Higgs particle whose low energy correction contributes to a cubic Yukawa type potential in position space.
Two long-ranged interactions consist of one repulsive tree level interaction of quadratic order $\mathcal{O}(q^{2}/\epsilon_{0})$, mediated by the scalar potential part $\Phi$ for massless degree of the U(1) gauge boson with the propagator \eqref{316} under the $R_{\xi}$ gauge fixing condition, and the other attractive tree level interaction of quadratic order $\mathcal{O}(g^{2})$, mediated by the gapless neutral scalar field of acoustic phonon with the propagator \eqref{322}. They are depicted by the following Feynman diagrams:

\begin{center}
\begin{tikzpicture}
\begin{feynman}[large]
\vertex (i1);
\vertex [below right=of i1] (a);
\vertex [below left=of a] (i2);
\vertex [right=of a] (b);
\vertex [above right=of b] (f1);
\vertex [below right=of b] (f2);

\diagram* {
(i2) -- [fermion] (a) -- [fermion] (i1),
(a) -- [photon] (b),
(f2) -- [fermion] (b) -- [fermion] (f1),
};
\end{feynman}
\end{tikzpicture}
\hspace{8em}
\begin{tikzpicture}
\begin{feynman}[large]
\vertex (i1);
\vertex [below right=of i1] (a);
\vertex [below left=of a] (i2);
\vertex [right=of a] (b);
\vertex [above right=of b] (f1);
\vertex [below right=of b] (f2);

\diagram* {
(i2) -- [fermion] (a) -- [fermion] (i1),
(a) -- [scalar] (b),
(f2) -- [fermion] (b) -- [fermion] (f1),
};
\end{feynman}
\end{tikzpicture}
\end{center}

%\begin{figure}[H]
%\centering
%\includegraphics[width=0.3\textwidth]{feynp4g1.png}
%\qquad
%\includegraphics[width=0.3\textwidth]{feynp4n1.png}
%\end{figure}

\noindent
In momentum space, the scattering amplitude of two complex scalar fields mediated by the scalar potential part of $\text{U}(1)$ gauge field is 
\begin{align}
V_{\Phi} (t, \boldsymbol{x}, t^{\prime}, \boldsymbol{x}^{\prime}, \omega_{\Lambda} ; q, c) =
\frac{q^{2}}{\epsilon_{0}} \int d\omega d^{2} \boldsymbol{k}\, 
dk_{z}e^{- i \omega (t - t^{\prime}) + i \boldsymbol{k} \cdot (\boldsymbol{x} - \boldsymbol{x}^{\prime}) + i k_{z} (z - z^{\prime})} 
\frac{1}{\frac{\omega^{2}}{c^{2}} 
- \boldsymbol{k}^{2}- k_{z}^{2}}
,\label{318}
\end{align} and that of mediated by the gapless neutral scalar field is
\begin{align}
V_{N} (t, \boldsymbol{x}, t^{\prime}, \boldsymbol{x}^{\prime}, \omega_{\Lambda} ; g, v_{N}) =
- g^{2}  \int d\omega d^{2} \boldsymbol{k}\, dk_{z}e^{-i\omega(t-t^{\prime}) +i \boldsymbol{k}\cdot (\boldsymbol{x}-\boldsymbol{x}^{\prime}) + i k_{z} (z-z^{\prime})} \frac{1}{ \frac{\omega^{2}}{v_{N}^{2}} -  \boldsymbol{k}^{2} - k_{z}^{2}}
.\label{323}
\end{align}
If the static approximation of negligible frequency parts, $\omega\sim0$, is again applied to \eqref{318}--\eqref{323} for consistency, the corresponding low energy interactions in position space for thick or thin superconducting samples consist of a 3- or 2-dimensional Coulomb potential from the massless degree of U(1) gauge field,  
\begin{align}
V_{\Phi}(\boldsymbol{x}, \boldsymbol{x}^{\prime}) 
= \begin{cases}
\frac{q^{2}}{4\pi \epsilon_{0}} \frac{1}{|\boldsymbol{x} 
- \boldsymbol{x}^{\prime}|} \qquad &\text{for $(1+3)$D}
\\ 
\frac{q^{2}}{2\pi \epsilon_{0}} 
\ln |\boldsymbol{x} - \boldsymbol{x}^{\prime}| \qquad 
&\text{for $(1+2)$D}
\\
\end{cases}
,\label{319}
\end{align}
and a Coulomb type potential from gapless neutral scalar field,
\begin{align}
V_{N}(\boldsymbol{x}, \boldsymbol{x}^{\prime}) = 
\begin{cases}
- \frac{g^{2}}{4\pi} \frac{1}{|\boldsymbol{x} - \boldsymbol{x}^{\prime}|} \qquad &\text{for $(1+3)$D}
\\ 
- \frac{g^{2}}{2\pi} \ln |\boldsymbol{x} 
- \boldsymbol{x}^{\prime}| \qquad &\text{for $(1+2)$D}
\\
\end{cases}
.\label{324}
\end{align}
Under this static approximation, the time-independent Coulomb potential 
\eqref{319} obtained in low energy limit implies that the origin of the mediated force is purely electric. As expected, the net interaction is given by the addition of a repulsive Coulomb interaction \eqref{319} mediated by the massless degree of U(1) gauge field and an attractive Coulomb type interaction \eqref{324} mediated by the gapless acoustic phonon.

Since these long ranged static forces do not involve any characteristic length scale, equal strengths of the forces are enough to cancel one another everywhere.
This long ranged net interaction without characteristic length scale vanishes everywhere if the cubic Yukawa type coupling $g$ has the critical coupling
\begin{align}
g_{\rm c} = \frac{q}{\sqrt{\epsilon_{0}}}
, 
\label{327}
\end{align}
whose value for the Cooper pair is $g_{\rm c} = - 1.08\times10^{-13} \, {\rm kg}^{1/2} {\rm m}^{3/2} {\rm s}^{-1}$ in tree level. Now, in the effective field theory of our consideration, the question why the electrostatic force between two Cooper pairs of charge $q=-2e$ disappears in superconducting phase is answered by zero net static force at least in tree level.
For weak cubic Yukawa type coupling $q/\sqrt{\epsilon_{0}}<g\le 0$, the strength of attractive phonon interaction is weaker than that of the repulsive Coulomb interaction and hence the net interaction between two Cooper pairs becomes repulsive everywhere. For strong cubic Yukawa type coupling $g>q/\sqrt{\epsilon_{0}}$, the strength of attractive phonon interaction is stronger than that of the repulsive Coulomb interaction and hence the net interaction between two Cooper pairs becomes attractive everywhere. In the vicinity of critical coupling \eqref{327}, the net interaction in static limit is feeble after the cancellation of repulsive electrostatic interaction and attractive phonon interaction.
Thus, the critical phonon coupling $g=g_{\rm c}$ provides the condition for superconductivity in this effective field theory.
If we apply this critical coupling \eqref{327} to the relation \eqref{348}, it gives a formula for the high cutoff frequency without free parameter,
\begin{align}
\omega_{\rm H} = \frac{2 m c v_{N} \xi}{\hbar \lambda_{\rm L}}
.
\end{align}
If the quantities in the right-hand side are already measured or calculated for a superconducting material, the value of high cutoff frequency $\omega_{\rm H}$ or equivalently the quartic self-interaction coupling of complex scalar field $\lambda$ is read in the given superconductor.

In the framework of perturbative quantum field theory, we show that attractive and repulsive static forces between two Cooper pairs are cancelled in the superconducting samples of the critical couplings $(\lambda,g)=(\lambda_{{\rm c}},g_{{\rm c}})$: First, electrostatic Coulomb force by the massless mode is perfectly cancelled by the phonon interaction in its static limit for the critical coupling \eqref{327} at least in leading tree level. Second, quantum corrections by both magnetostatic force by the massive mode of $\text{U}(1)$ gauge field and the massive Higgs interaction vanishes for the critical coupling \eqref{312} at least in leading 1-loop level.
Different from the known limit of critical quartic self-interaction coupling of complex scalar field $\lambda=\lambda_{{\rm c}}$ \cite{Abrikosov:1956sx}, the limit defined by critical cubic Yukawa type coupling \eqref{327} is new and affects dominantly in long range due to power law behavior, the electric field of $1/r^{2}$ in three spatial dimension or $1/r$ in two spatial dimension.
Thus, in the limit of critical coupling \eqref{327} of the effective field theory of the proposed action \eqref{201}, slowly moving Cooper pairs can behave almost noninteracting despite of their charges because of cancellation of the repulsive Coulomb interaction and attractive phonon interaction. If they are distant, such almost noninteracting character is unaffected by the other quartic self-interaction coupling $\lambda$ which gives short ranged interaction and hence is irrespective of the types of superconductors.

Every cancellation of the interactions discussed above is based on the static approximation valid only for extremely low frequency limit. It means that inclusion of the contributions of all frequencies higher than the low critical frequency induces discrepancy between the attractive and repulsive interactions even at the critical coupling $g=g_{\rm c}$. A naive way to test with ignoring complicated issues, e.g., renormalization, is to introduce a cutoff frequency $\omega_{\Lambda}$ by hand and compute the discrepancy for a given pair of couplings $(\lambda,g)$, which is defined,
\begin{align}
\Delta = \bigg| \frac{V_{\Phi} (t, \boldsymbol{x}, t^{\prime}, \boldsymbol{x}^{\prime}, \omega_{\Lambda} ; q, c) + V_{N} (t, \boldsymbol{x}, t^{\prime}, \boldsymbol{x}^{\prime}, \omega_{\Lambda} ; g, v_{N})}{V_{\Phi} (t, \boldsymbol{x}, t^{\prime}, \boldsymbol{x}^{\prime}, \omega_{\Lambda} ; q, c) - V_{N} (t, \boldsymbol{x}, t^{\prime}, \boldsymbol{x}^{\prime}, \omega_{\Lambda} ; g, v_{N})} \bigg|
.\label{403}
\end{align}
Note that, for the critical temperature of conventional superconducting samples $T_{{\rm c}}\approx 4{\rm K}$, the cutoff frequency is of the order $\omega_{\Lambda} \sim 10^{11} \text{Hz}$.
If we consider an elapsed time comparable to the cutoff frequency, $t - t^{\prime} \approx 2\pi/\omega_{\Lambda}$, with negligibly small distance $d\approx 0$, the discrepancy \eqref{403} is given by
\begin{align}
\Delta \approx \left| \frac{1 - (\frac{v_{N}}{c})^{2} \frac{q^{2}/\epsilon_{0}}{g^{2}} }{1 + (\frac{v_{N}}{c})^{2} \frac{q^{2}/\epsilon_{0}}{g^{2}}} \right| \sim \left| \frac{1 - 10^{-11} \big( \frac{g_{\rm c}}{g} \big)^{2} }{1 + 10^{-11} \big( \frac{g_{\rm c}}{g} \big)^{2} } \right|
,\label{402}
\end{align}
where the two propagation speeds of light $c = 3.0\times 10^{8} \text{m/s}$ and phonon $v_{N} \sim 10^{3} \text{m/s}$ are used.
The approximated discrepancy \eqref{402} shows large discrepancy $\Delta \approx 1$ for almost all the range of phonon coupling except the negligibly tiny cases of $g \simeq 10^{- 5 \sim - 6} g_{\rm c}$.
Oppositely, we consider zero elapsed time $t - t^{\prime} \approx 0$ and a finite distance $d = \sqrt{(\boldsymbol{x} - \boldsymbol{x}^{\prime})^{2} + (z - z^{\prime})^{2}}$, the discrepancy $\Delta$ becomes a function of the distance $d$ and cutoff $\omega_{\Lambda}$,
\begin{align}
\Delta \approx \left| \frac{\big[1 - (\frac{g}{g_{\rm c}})^{2}\big] - \big\{ \cos (\frac{\omega_{\Lambda}}{c} d) - (\frac{g}{g_{\rm c}})^{2} \cos \big[ (\frac{c}{v_{N}}) \frac{\omega_{\Lambda}}{c} d \big] \big\} }{\big[1 - (\frac{g}{g_{\rm c}})^{2}\big] - \big\{ \cos (\frac{\omega_{\Lambda}}{c} d) + (\frac{g}{g_{\rm c}})^{2} \cos \big[ (\frac{c}{v_{N}}) \frac{\omega_{\Lambda}}{c} d \big] \big\}} \right|
,\label{404}
\end{align}
where two extremely different frequencies $\omega_{\Lambda}$ and $(c/v_{N}) \omega_{\Lambda} \sim 10^{-5} \omega_{\Lambda}$ are included.
Even at the critical coupling \eqref{327}, it has nonzero value for most of distances due to interference of two cosine terms of completely different frequencies.
Conventional superconductivity is generically sensitive to temperature.
Once higher frequency modes are participated in contribution to the discrepancy \eqref{403}, the static force balance at the critical coupling is easily collapsed.
The noticed discrepancy appearing away from the static limit can reconcile comprehensively and quantitatively the character why superconductivity, at least conventional superconductivity based on $s$-wave, is fragile even above the very low critical temperature.
It is intriguing to devise a simple and quantitative procedure for evaluation of the critical temperature of each superconducting material in the framework of effective field theory in addition to the accurate theoretical evaluation scheme to determine the maximum critical temperature \cite{McMillan:1968pr, Allen:1968pr, Giustino:1968pr}.

Throughout this section, we obtain a new critical cubic Yukawa type coupling \eqref{327} in addition to that of the quartic self-interaction coupling \eqref{312} by perturbative analysis in the previous section.
At the critical coupling $\lambda=\lambda_{{\rm c}}$ \eqref{312}, the correlation length $\xi$ \eqref{303} and the London penetration depth $\lambda_{\rm L}$ \eqref{302} become equal with unit Ginzburg-Landau parameter \eqref{314}. Therefore, experimentally, superconducting materials are classified into two, type I in weak coupling regime $0 \le \lambda/\lambda_{\rm c} < 1$ and type I$\!$I in strong coupling regime $\lambda/\lambda_{\rm c} > 1$, by this borderline of critical coupling. Quantum field theoretic calculation shows that this critical value is understood as an interaction balance of the 1-loop level attractive interaction mediated by the massive Higgs boson and the 1-loop level repulsive interaction mediated by the massive degree of U(1) gauge boson under static approximation. Another noteworthy tree level cancellation occurs at the critical coupling $g=g_{{\rm c}}$ \eqref{327} of the cubic Yukawa type interaction between phonon and Cooper pair. The attractive long ranged interaction mediated by the gapless acoustic phonon exactly cancels the repulsive Coulombic interaction mediated by the massless degree of U(1) gauge boson.
This perfect cancellation can explain almost free motion of charged Cooper pairs in superconducting phase.
This new critical phonon coupling gives a condition for superconductivity $g \simeq g_{\rm c}$.
The interaction balance only in static limit may explain temperature-fragile sensitivity of conventional superconductivity in accordance with two quite different characteristic speeds of phonon and photon, $v_{\rm N}/c\sim10^{-5} \ll 1$.

Though we explain the temperature-fragile character of superconductivity by introducing a cutoff frequency in tree level, straightforward application to the interactions by the 1-loop loops as in \eqref{360}--\eqref{361} does not work due to unavoidable regularization procedure in the middle of loop calculation.
However, despite of the lack of specific calculations, it seems likely that the propagators of qualitatively different $\omega$ behaviors in \eqref{360} and \eqref{361} may result in an intriguing suggestion for experiments.
For example, the borderline between type I and I$\!$I in a given superconducting sample can possibly become fuzzy by temperature change even below the critical temperature. Thus, the lower the temperature is for a given superconducting material, the more accurate the value of the quartic self-interaction of complex scalar coupling $\lambda$ can be measured. 

\section{Conclusion and Discussion}

In this paper, a field-theoretic description of conventional superconductivity is proposed. The action of effective field theory \eqref{201} consists of a Schr\"odinger type complex scalar field of Cooper pair, a U(1) gauge field of electromagnetism, and a nonrelativistic neutral scalar field of acoustic phonon with constant background charge density.
Their interactions involve a quartic self-interaction of complex scalar field and a cubic Yukawa type interaction between neutral and complex scalar fields in addition to electromagnetic interaction through minimal coupling. Counting of canonical dimension of the fields says that the proposed effective field theory is renormalizable  and hence this renormalizability is worth tackling in the framework of 
 perturbative quantum field theory.

The phonon field participates in our effective field theory even after the Cooper pair, a composite of two electrons, appears as a composite complex scalar field. We explain in the context of effective field theory the possible reason why such residual gapless phonon survives at low energy while its high frequency modes are integrated out to produce quartic self-interaction of complex scalar field. This plausible argument must be confirmed through systematic approach in the scheme of effective field theory.

We begin with the equilibrium of zero vacuum expectation value of the phonon field and find the superconducting vacuum of the complex scalar field with nonzero vacuum expectation value identified as the square root of constant matter density of background charge. This energetically favored constant vacuum configuration explains zero electrical resistance and perfect diamagnetism, e.g., the Meissner effect in the presence of constant external magnetic field is understood as the recovery of 
symmetry-broken vacuum to minimize the energy. Since the superconducting vacuum is obtained in the topologically trivial sector, it is worth studying the sectors of nontrivial first homotopy manifested by topological 
vortices~\cite{FTSC} which are compared to the Abrikosov-Nielsen-Olesen vortices~\cite{Abrikosov:1956sx}.

We newly find a critical coupling of the cubic Yukawa type interaction between neutral and complex scalar field \eqref{327}, proportional to the charge of a Cooper pair $q=-2e$. When the long ranged Coulombic repulsion mediated by the massless mode of U(1) gauge field is exactly cancelled by the long ranged  attraction mediated by the gapless acoustic phonon, this critical phonon coupling is obtained  in leading order with taking static limit. Disappearance of the net long ranged interaction between two Cooper pairs explains at least in tree level that the reason why two Cooper pairs are almost noninteracting despite of strong Coulomb interaction at the critical phonon coupling and their interaction becomes feeble in the vicinity of the critical coupling. Since the interaction balance holds strictly only in static limit, quite different characteristic speeds of photon and phonon of the order of $10^{-5}$ can induce large discrepancy from the interaction balance away from the static limit, e.g., temperature. Thus this discrepancy provides a plausible explanation of the reason why conventional superconductivity of $s$-wave is so fragile to growing temperature and has such low critical temperature. Approach by use of Matsubara frequencies in temperature field theory may be beneficial for concrete studies by use of the effective field theory of consideration \cite{Kapusta:1989tk}.

The critical coupling of the quartic self-interaction of complex scalar field \eqref{312} was introduced in the beginning stage of the Ginzburg-Landau theory through equality of the London penetration depth and the correlation length \cite{Abrikosov:1956sx}. In this work, it is understood as another static interaction balance between the repulsion mediated by the massive degree of virtual photon and the attraction mediated by massive neutral Higgs at least in 1-loop level. For weak coupling regime of longer correlation length, the strength of attractive Higgs interaction becomes weaker. Then, in type I superconductors, the net short ranged interaction is repulsive in shorter distance but turns to be attractive at relatively larger distance. Contrarily, for strong coupling regime of shorter correlation length, the strength of attractive Higgs interaction becomes stronger. Then, in type I$\!$I superconductors, the net short ranged interaction is attractive in shorter distance but turns to be repulsive at relatively larger distance. Now the obtained critical value of cubic Yukawa type coupling $g_{{\rm c}}$ \eqref{327} provides  the condition for conventional superconductivity possessing commonly the superconducting vacuum. Superconductors obtained in the effective field theory of consideration are divided into type I and I$\!$I by the borderline of the critical value of the quartic self-interaction coupling 
$\lambda_{{\rm c}}$ \eqref{312} as obtained first in the Ginzburg-Landau theory.  Since our analysis is made in the leading tree level for the cubic Yukawa type coupling and in 1-loop level for the quartic self-interaction coupling by perturbative method, higher order calculations including more loops are worth performing in perturbative regime, including propagator and vertex corrections.
It may be intriguing to find the equality of these two critical couplings protected even after further quantum corrections, that implies consistency with nonperturbative analysis by use of topological vortices \cite{FTSC}.

According to field theoretic study by use of the proposed action \eqref{201}, some key properties of superconductivity of $s$-wave are figured out in the context of perturbative quantum field theory. Once this effective field theory is established as the quantum many body description of conventional superconductors after passing indispensable tests, the known and new physical quantities including the time-dependent quantum processes can be calculated by tractable systematic methods and then await experimental data.

\section*{Acknowledgement}

The authors would like appreciate Kwang-Yong Choi, Chanyong Hwang, Chanju Kim, O-Kab Kwon, Hyunwoo Lee, Tae-Ho Park, and  D. D. Tolla for discussions on various topics of condensed matter physics and field theory. This work was supported by the National Research Foundation of Korea(NRF) grant with grant number NRF-2022R1F1A1073053 and RS-2019-NR040081 (Y.K.).

\end{document}